\documentclass[aps,prl,twocolumn,groupedaddress,floatfix,amsmath,amssymb]{revtex4}
\usepackage{graphicx}
\usepackage{bm}
\begin{document}

\title{Spin-Resolved Self-Doping Tunes the Intrinsic Half-Metallicity of AlN Nanoribbons}

\author{Alejandro Lopez-Bezanilla$^{1,2}$}
\email[]{alejandrolb@gmail.com}
\author{P. Ganesh} 
\email[]{ganeshp@ornl.gov}
\author{P. R. C. Kent}
\author{Bobby G. Sumpter}
\affiliation{$^{1}$Oak Ridge National Laboratory, One Bethel Valley Road, Oak Ridge, Tennessee, 37831-6493, USA}
\affiliation{$^{2}$Materials Science Division, Argonne National Laboratory, 9700 S. Cass Avenue, Lemont, IL 60439, USA}

\begin{abstract}
We present a first-principles theoretical study of electric field-and strain-controlled intrinsic half-metallic properties of zigzagged aluminium nitride (AlN) nanoribbons. We show that the half-metallic property of AlN ribbons can undergo a transition into fully-metallic or semiconducting behavior with application of an electric field or uniaxial strain. An external transverse electric field induces a full charge screening that renders the material semiconducting. In contrast, as uniaxial strain varies from compressive to tensile, a spin-resolved selective self-doping increases the half-metallic character of the ribbons. The relevant strain-induced changes in electronic properties arise from band structure modifications at the Fermi level as a consequence of a spin-polarized charge transfer between $\pi$-orbitals of the N and Al edge atoms in a spin-resolved self-doping process. This band structure tunability indicates the possibility ofdesigning magnetic nanoribbons with tunable electronic structure by deriving edge states from elements with sufficiently different localization properties. Finite temperature molecular dynamics reveal a thermally stable half-metallic nanoribbon up to room temperature.

\end{abstract}

\maketitle

\section{Introduction}
Encouraging theoretical studies and experimental 
achievements in the preparation and characterization
of new materials have increased the scientific efforts to 
find alternative materials to those used in conventional
charge-based electronic devices, optical switches and 
spin-based electronics [1]. A material exhibiting metallic 
behavior for electron spins with one orientation and 
insulating for the spins with the opposite orientation
is highly desirable in spintronics. Heusler alloys and 
compounds based on rare earths are among the 
prominent half-metallic materials able to conduct 
electrons in spin-polarized currents [2, 3]. Recently, 
the versatility in shape and compositions of newly 
synthesized monolayered materials such as graphene 
or MoS$_2$ [4] allow us to envision new devices with half- 
metallic features. External fields have shown promise
in tuning the electronic properties of nanoribbons [5],
as demonstrated in graphene nanoribbons where
half-metallicity may be realized [6]. Also, a scheme 
based on the oxidation of zigzagged graphene 
nanoribbons combined with the application of an 
external electric field allowed for the prediction of 
half-metallicity in a carbon-based material [7]. In 
boron nitride, it has been reported that strain fields 
derived from axial deformation of nanoribbons could 
reduce the large electronic band gap [8]. These studies 
offer an interesting opportunity of individually con-
trolling the response of the two edges of a nanoribbon 
to achieve wider tunability of the electronic structure.

Aluminium nitride (AlN) is a large band gap 
semiconductor with extraordinary physical properties 
such as small thermal expansion coefficient, high 
thermal conductivity and a reasonable thermal match 
to Si and GaAs [9]. AlN also has attractive piezoelectric 
properties which may be suitable for surface acoustic 
wave device applications [10]. Balasubramanian et al. 
provided evidence that tubular AlN structures composed of hexagonal rings of Al and N atoms adopting 
sp$^2$ hybridization can be fabricated [11]. Zhukovskii 
et al. reported the structural and electronic properties 
of single-walled AlN nanotubes of different chiralities 
and sizes [12]. The successful experimental synthesis 
and characterization by Xie et al. [13] of pure crystalline 
AlN nanoribbons with various edge morphologies 
suggests that controlled growth of desirable architectures based on this group III nitride material may be 
possible.

\begin{figure}[htp]
 \centering
 \includegraphics[width=0.5 \textwidth]{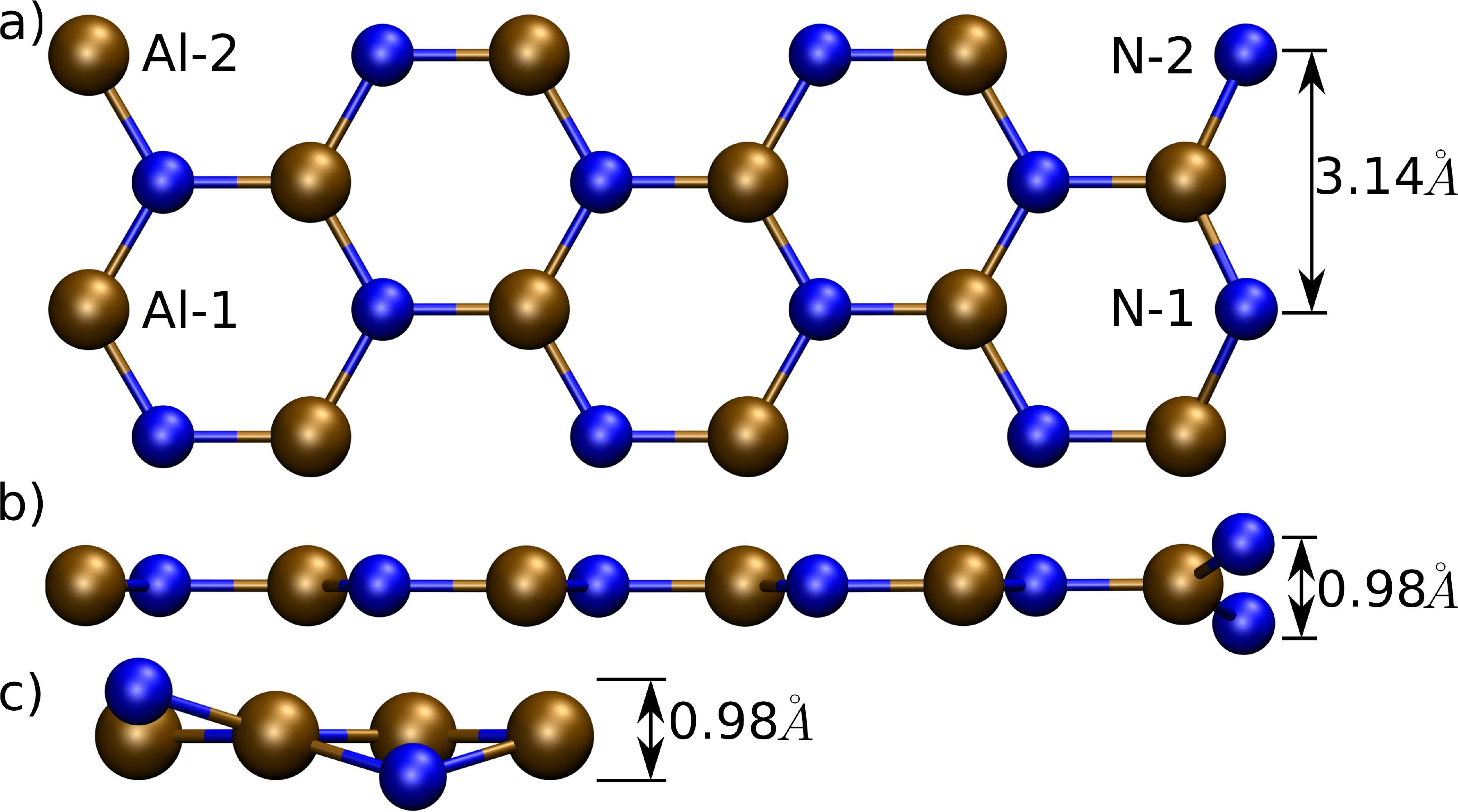}
 \caption{Different perspectives of the unit cell geometry of the 
fully relaxed [AlN]$_6$(2) ribbon. (a) Top view of the hexagonal 
network with a unit cell parameter of 2 \AA\ $\times$ 3.14 \AA. Edge atoms 
are labeled as Al-1, Al-2, N-1 and N-2. (b) Perspective along the 
ribbon axis showing N atoms displaced above and below the 
ribbon plane a relative distance of 0.98 \AA\ to each other. In (c) this 
displacement is also shown in a side view of the N edge. Brown 
and blue spheres represent Al and N atoms, respectively.}                                                                                                                                                                                                                                                                                                                                             
 \label{fig1}
\end{figure}

In this paper we present theoretical evidence of the 
tunability of half-metallic and magnetic properties of 
pristine aluminum nitride nanoribbons by means of 
both an external applied electric field and longitudinal
(axial) strain. A broad spectrum of electronic configurations ranging from a half-metallic to an insulating 
regime are demonstrated, indicating that purely 
physically oriented approaches are effective methods 
to control the electric and magnetic behavior without 
chemical functionalization or doping. Unlike results 
reported for graphene [6] and boron nitride [8] nano-
structures, but similar to the behavior of ultrathin 
metal oxide films [14], zigzag pristine AlN nanoribbons 
exhibit spin-sensitive response to axial strain, yielding
a spin-resolved self-doping process that depends on
both the spin-configuration of the ribbon and the
external stimulus. Fundamentally, we rationalize this
tunability as having its origin in the opposing edge 
states that are derived from orbitals with sufficiently 
different localization. 

\section{Computational Methodology}

The geometry optimizations and electronic structure 
calculations were performed by means of the SIESTA 
density functional theory (DFT) based code [15,16]. For each ribbon structure we tested 
multiple initial magnetic configurations, particularly 
along the nanoribbon edges, to determine their relative 
energies. We used a double-polarized basis set within the spin-dependent 
general gradient approximation (PBE). AlN nanoribbons 
were modeled within a supercell with at least 10 \AA\ of vacuum 
to avoid interactions between neighboring cells. Atomic 
positions were relaxed with a force tolerance of 5 meV/\AA. 
The Brillouin zone integration used a Monkhorst-Pack
sampling of 1 $\times$ 1 $\times$ 32 k-points for two-row ribbons. The 
radial extension of the localized orbitals used a kinetic energy
cutoff of 70 meV. The numerical integrals are computed on 
a real space grid with an equivalent cutoff of 500 Ry. 
 
Selected Perdew-Burke-Ernzerhof (PBE) results from 
SIESTA were also validated with the plane wave 
approach VASP [18, 19] to guarantee the robustness 
of the results. A plane wave cut-off of 400 eV and Brillouin zone sampling 
identical to the SIESTA calculations was used. Electronic 
states were occupied using either the Methfesel-Paxton 
method or Gaussian smearing of up to 0.25 eV. Molecular 
dynamics simulations were performed in the NVT ensemble 
for a 4-unit super-cell using only the $\Gamma$-point, starting from 
the ground-state spin-structure for 0\% and 8\% strain at T = 
150 K and T = 300 K. 2.5 ps long runs were performed after 
an initial equilibration time of 0.8 ps with a 1 fs time-step.

\section{Results and Discussion}

\begin{figure*}[htp]
 \centering
 \includegraphics[width=0.95 \textwidth]{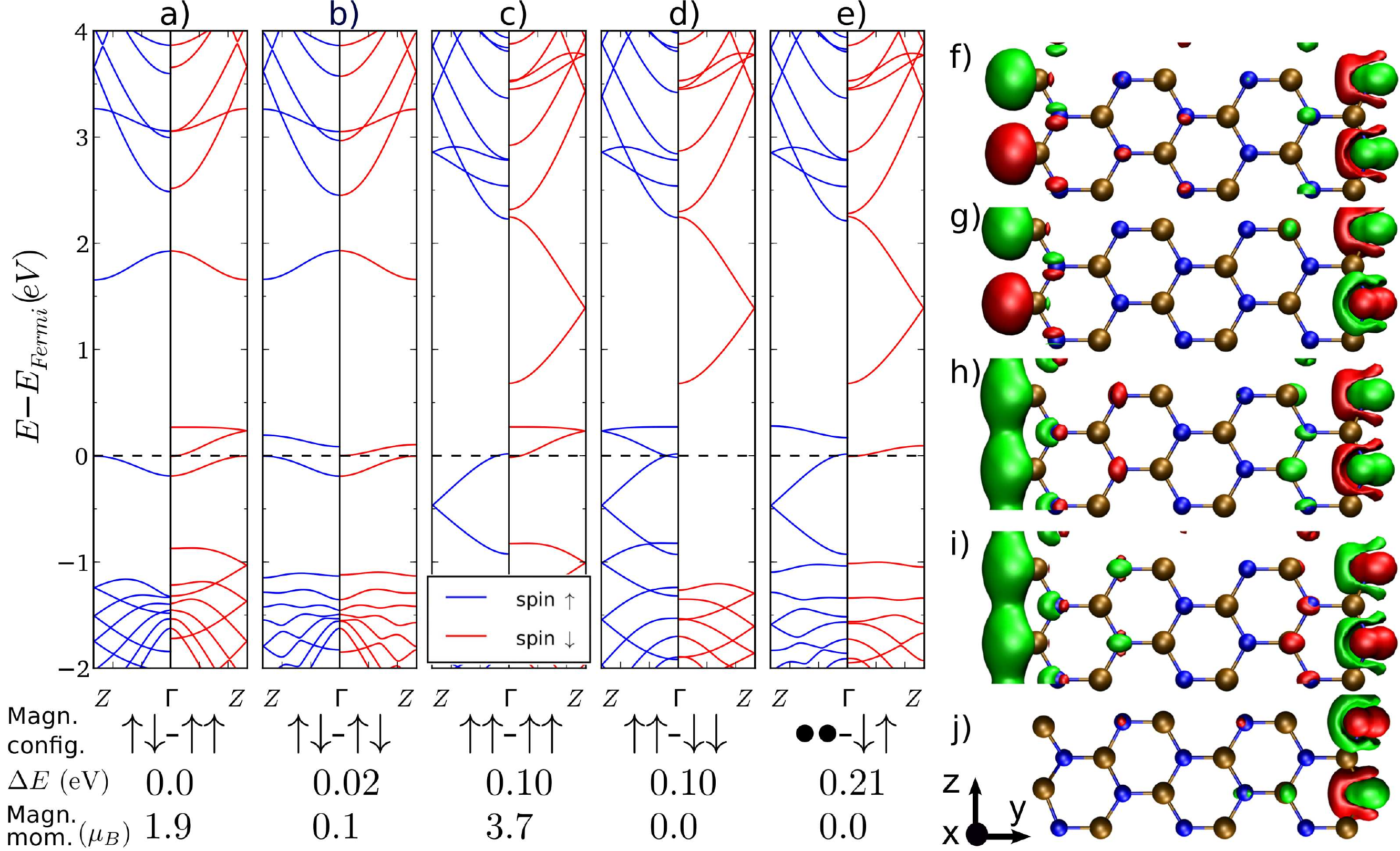}
 \caption{Figure \ref{fig2} From (a) to (e), in energetically increasing order, the electronic band structure of the five [AlN]$_6$(2) ribbon configurations 
studied in this paper. The spin orientations for Al- and -N edge atom are given by arrows, the relative formation energies ($\Delta$E) and the 
magnetic moment ($\mu_B$). Horizontal dashed lines indicate the Fermi level. From (f) to (j), the real space distribution of the net-spin 
density corresponding to the electronic states of [AlN]$_6$(2) in panels from (a) to (e). Isosurfaces are plotted at 10$^{-3}$ e/\AA$^3$
green isosurfaces correspond to net spin-$\uparrow$ and spin-$\downarrow$ electron densities, respectively.}                                                                                                                                                                                                                                                                                                                                             
 \label{fig2}
\end{figure*} 

In the following we refer to a bare-edge zigzagged 
AlN-NR composed of n rows, each with m dimer lines, 
as [AlN]$_m$(n). The non-passivated dangling sp$^2$ $\sigma$-bonds 
of the edge atoms are responsible for the unique 
magnetic and electronic features of the ribbons, so our 
ribbons are sufficiently wide to avoid interactions
between these edge states. Figure \ref{fig1} shows a schematic
representation of a 1.43 nm wide [AlN]$_6$(2) ribbon. A 
striking characteristic of this nanoribbon is that, in 
contrast to the Al zigzagged edge that exhibits a flat
geometry, N atoms on the opposite edge present a
zigzagged ordering above and below the ribbon plane.

\begin{figure*}[htp]
 \centering
 \includegraphics[width=0.5 \textwidth]{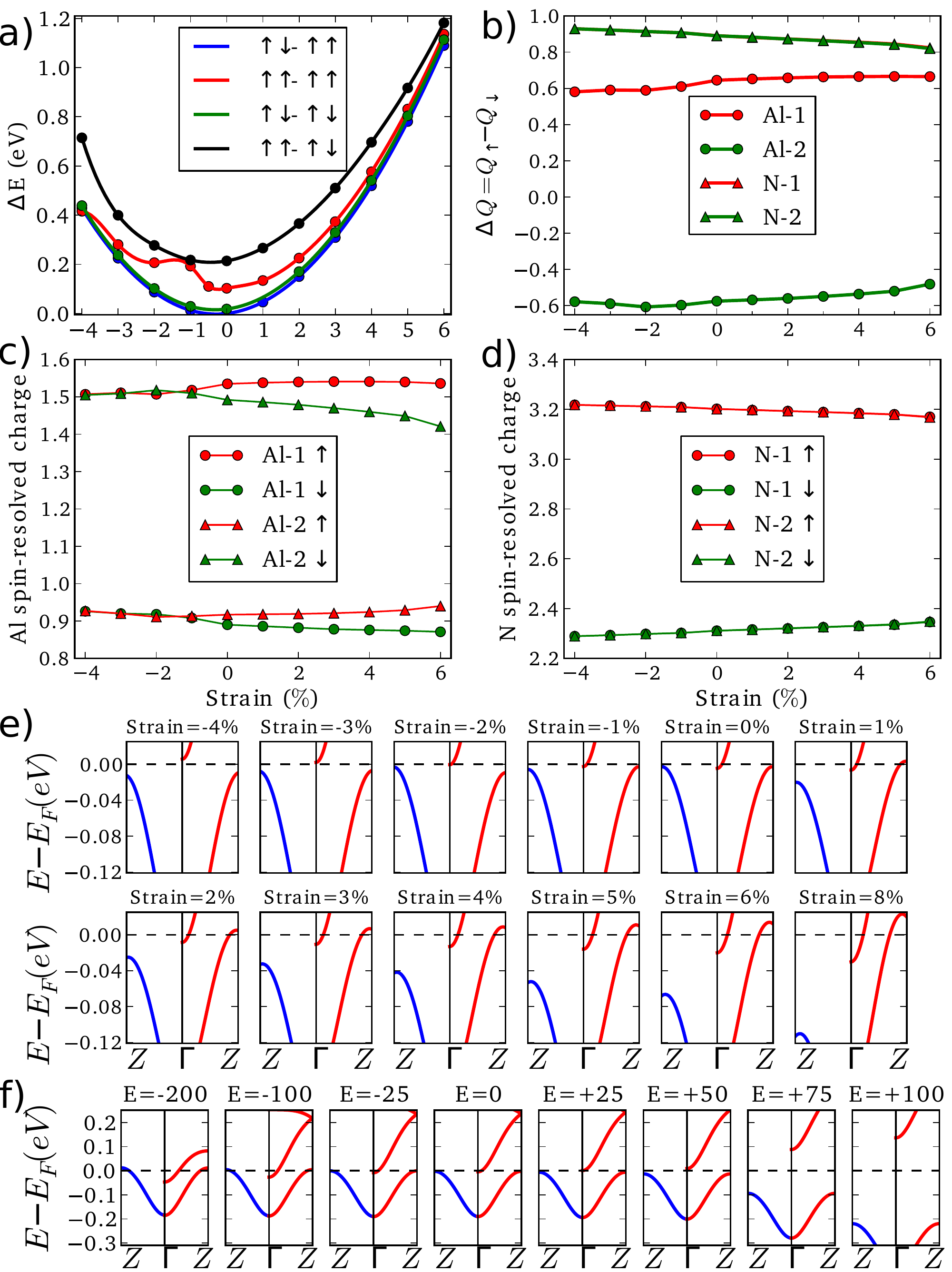}
 \caption{ (a) Computed energy vs. strain curves of a [AlN]$_6$(2) ribbons in different spin configurations. (b) Total spin polarization (Q$_\uparrow$ -
Q$_\downarrow$) of edge atoms following the nomenclature of Figure \ref{fig1}. The total amount of spin-resolved charge in the edge Al and N atoms are shown 
in (c) and (d), respectively, for different degrees of uniaxial strain. In (e) the spin-resolved band structure, within a narrow energy window 
around the Fermi level, for different unit cell parameters is shown. (f) Evolution of the spin-resolved band structure for different external
electric fields applied transversally to the $\uparrow\downarrow $-$\uparrow\uparrow$ configuration. The initially metallic system for E = -200 meV turns into half-metallic 
for E = -25 meV/\AA\ and semiconducting for E = 50 meV/\AA\ on. The electric field is positive for the Al to N sense.}                                                                                                                                                                                                                                                                                                                                             
 \label{fig3}
\end{figure*}

The vertical distance between N edge atoms is 0.98 \AA\ (see Figures. 1(b) and 1(c)). This allows the edge Al-N 
bond length to be the same as the inner region of the 
ribbon, namely, 1.81 \AA. This bond length is slightly 
enlarged for bonds at the Al edge. All ribbons 
considered in this paper are free-standing, as we are 
interested in understanding their intrinsic properties. 
Both their geometry and electronic properties could 
undergo variations depending on a support substrate.
We now scrutinize the electronic properties of the
[AlN]$_6$(2) ribbon by performing an analysis on the
arrangement of five non-equivalent spin configurations 
shown in Figure \ref{fig2}. Only the $\uparrow\uparrow $-$\uparrow\downarrow$ configuration deviates
from the initial spin arrangement to end in the $\bullet\bullet$-$\uparrow\uparrow$ 
configuration, where $\bullet\bullet$ signifies zero magnetic 
moment. A comparison of the energy and the mag-
netic moments of each configuration is provided in 
Figure \ref{fig2}. The ground state corresponds to a system with 
total magnetic moment of 1.9 $\mu_B$ and an antiferro-
magnetic spin coupling at the Al edge, whereas the
opposite N edge exhibits a ferromagnetic coupling 
between neighboring displaced N atoms (see Figure \ref{fig2}(f)).
The highest occupied states (HOS) are below the 
Fermi level, and the lowest unoccupied state (LUS) of 
the spin-down channel crosses this limit, yielding a 
ground state of the [AlN]$_6$(2) ribbon with half-metallic 
semi-metal features. All other configurations are 
characterized by different spin at the ribbon edges 
with a strongly localized character. It is worth noting 
that flipping the spin of one of the N edge atoms is 
energetically more expensive than aligning the spins 
in a ferromagnetic ordering at the Al edge. In this case 
two energetically equivalent configurations, $\uparrow\uparrow $-$\uparrow\uparrow$ and 
$\uparrow\uparrow $-$\downarrow\downarrow$, are obtained with all spins ferromagnetically 
coupled along both edges, differing only in their 
relative orientation across the ribbon width. The 
electronic band diagram of the $\uparrow\uparrow $-$\downarrow\downarrow$ configuration
plotted in Figure \ref{fig2}(d) reveals that two electronic states 
cross the Fermi level for the spin-up channel, but there 
is a band gap of $\sim$2 eV for the spin-down channel.
This particular spin arrangement at the edges of the 
ribbon yields a half-metallic non-magnetic [AlN]$_6$(2)
ribbon, 100 meV above the ground state. To gain
further insight into the localized character of these
magnetic states at the ribbon edges, we performed a 
projected density of states analysis to determine that 
the main contributions to the electronic states near the 
Fermi level comes from the $\pi$-orbitals of the edge atoms. 
Specifically, the py and s-orbitals of Al-edged atoms 
have a large contribution to the HOS whereas the 
LUS is entirely formed from the py and the px orbitals 
of the N-edged atoms, with a small contribution from 
the s-orbitals of the same atoms.

The piezoelectric properties of AlN already motivate 
its integration with Si integrated circuits [17]. Given 
the current ability to transfer AlN thin-films onto 
many different substrates, it is reasonable to explore 
the possible changes in the electronic and magnetic
response to applied strain on AlN nanoribbons. We 
study the effect of axial deformations on the electronic
states and magnetic properties of [AlN]$_6$(2) ribbon.
In Figure \ref{fig3}(a) the computed energy vs. strain curves 
for a [AlN]$_6$(2) ribbon are plotted for the five non- 
equivalent spin configurations. For each system the 
energy evolves smoothly, and there is no switching in 
the energetic ordering of the different spin arrange
ments for the considered lattice parameter variation 
range, from -4\% up to 6\%, in the z direction (see Figure \ref{fig2}). 
Only one value of the $\uparrow\uparrow $-$\downarrow\downarrow$ configuration diverges
from the general quadratic trend exhibited by all 
configurations upon stretching and compression of the 
unit cell parameter. Closer examination of the atomic 
displacements with compressive strain suggests a small
barrier to local atomic distortion, similar to isostructural 
phase transitions in other layered materials [20]. 

In the band diagrams of Figure \ref{fig3}(e) we show the 
evolution of the electronic states, near the Fermi level, 
as a function of the lattice parameter variation fo the 
most relevant configuration. For compression values 
of -4\% and -3\% of the $\uparrow\downarrow $-$\uparrow\uparrow$ configuration, the system 
develops small band gap for both spin channels that 
renders the [AlN]$_6$(2) ribbon semiconducting. With
increasing strain one of the spin channels remains 
metallic while the HOS of the other spin channel
shifts down from the Fermi level, increasing the half-
metallicity. The total magnetic moment saturates to
2 $\mu_B$. A better understanding of the origin of the
half-metallicity enhancement with increasing strain
can be obtained from an analysis of the Mulliken
population of the edge atoms. The evolution of the
charge in each edge atom as a functional of the applied 
strain is plotted in Figure \ref{fig3}(b), showing the trend of Al 
edge atoms to gain some charge (N atoms to lose) as 
the ribbon increases the cell vector. Examining each 
spin channel separately in Figures 3(c) and 3(d), one 
notices that a net spin-up charge transfer takes place 
from the N edge to the Al edge whereas spin-down 
flows in the opposite direction, from the Al edge to 
the N edge. The Al to N charge transfer process is 
visible in the spin-resolved energy band diagrams of 
Figure \ref{fig3}(e): On increasing the strain starting with a 
strain of -4\%, the spin-down band below Fermi, that 
is mostly derived from the Al $\pi$-orbitals, shifts up and 
crosses the Fermi level as it is emptied. Simultaneously 
the spin-down band above the Fermi level that is 
derived mainly from the N $\pi$-orbitals on the opposite 
edge shifts down crossing the Fermi level, rendering the 
system half-metallic. Overall, a net charge migration 
from the N edge to the opposite Al edge occurs with 
increasing strain, in a spin-selective self-doping process 
between the two edges. 

The analysis presented above suggests that the 
magnetic polarization of the ribbon with electronic 
bands derived from atomic orbitals at opposite edges 
with sufficiently different electron localization pro
perties and with sufficient local spin densities of both 
channels on the more metallic edge, is the origin of the 
enhanced half-metallicity. To verify this assumption 
we have calculated the energy band diagrams of the 
ground state of a [AlN]$_6$(2) ribbon ($\uparrow\downarrow $-$\uparrow\uparrow$) for different 
external transversally applied electric fields (i.e., from 
Al edge to the N edge at positive fields using a 
sawtooth potential), in order to induce a charge flow 
that screens the external field across the ribbon width 
without distinguishing between spins. In Figure \ref{fig3}(f) we 
plot the spin resolved band structure as a function of 
a field intensity ranging from E = -200 meV/\AA\  up to 
E = +100 meV/\AA. The general behavior observed for 
the Al $\pi$-orbital states is a symmetric response to the 
external field, decreasing their energy relative to the 
Fermi level simultaneously for all E. Subsequently, 
the half-metallicity is barely modified for E values 
close to 0, and for most of the E values the ribbon 
exhibits a metallic (from E = -200 up to E = -50 meV/\AA) 
or semiconducting (from E = +50 meV/\AA\  up to E = 
+100 meV/\AA) behavior.

\begin{figure}[htp]
 \centering
 \includegraphics[width=0.5 \textwidth]{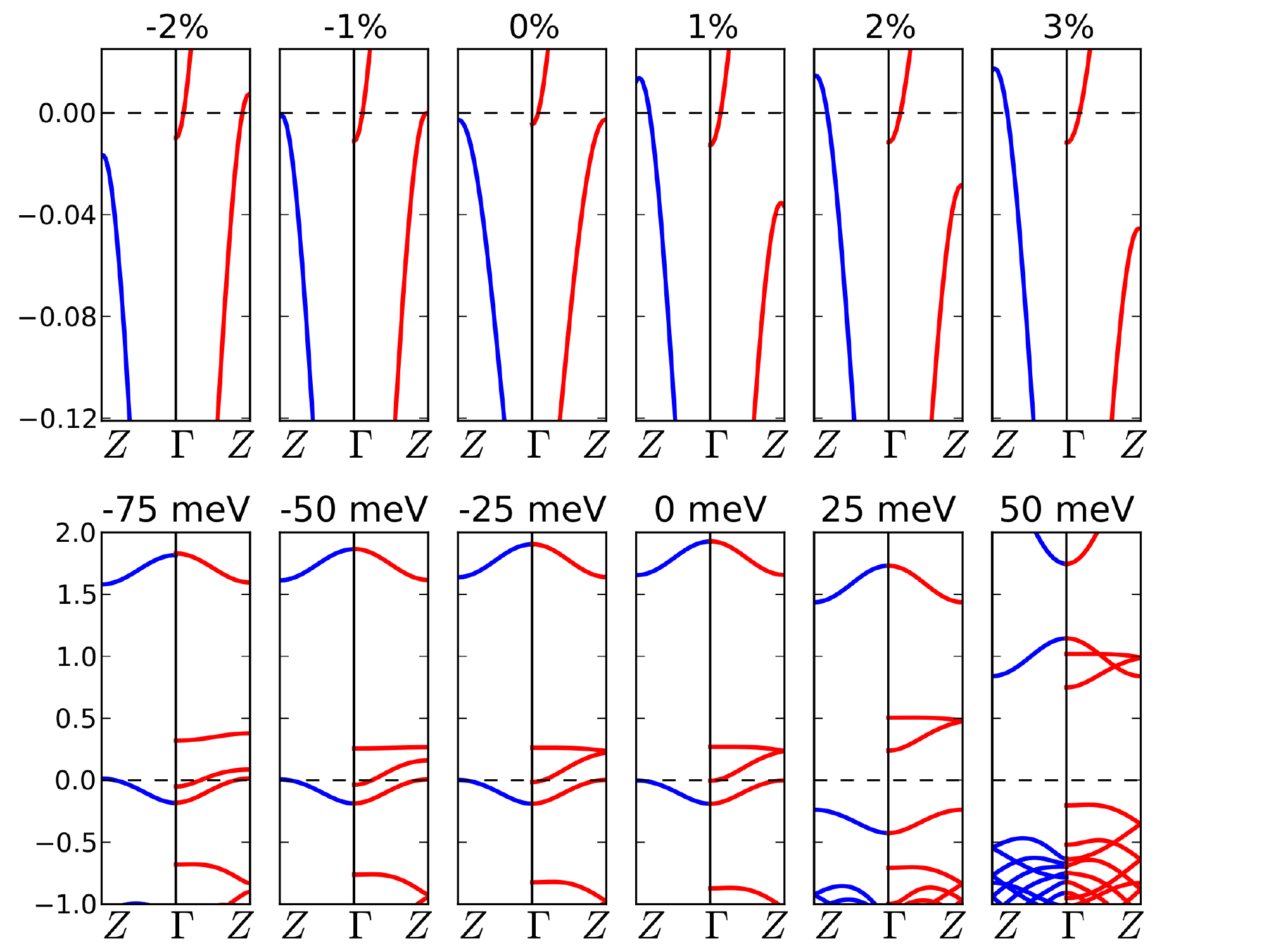}
 \caption{The tunability of the [AlN]$_{18}$(2) ribbon. Upper panels show the evolution of the ribbon from half-metallic to metallic from 
compressive to stretching unit cell values. Lower panels show the evolution from metallic to semiconducting behavior for different 
electric fields.}                                                                                                                                                                                                                                                                                                                                             
 \label{fig4}
\end{figure}

\begin{figure*}[htp]
 \centering
 \includegraphics[width=0.95 \textwidth]{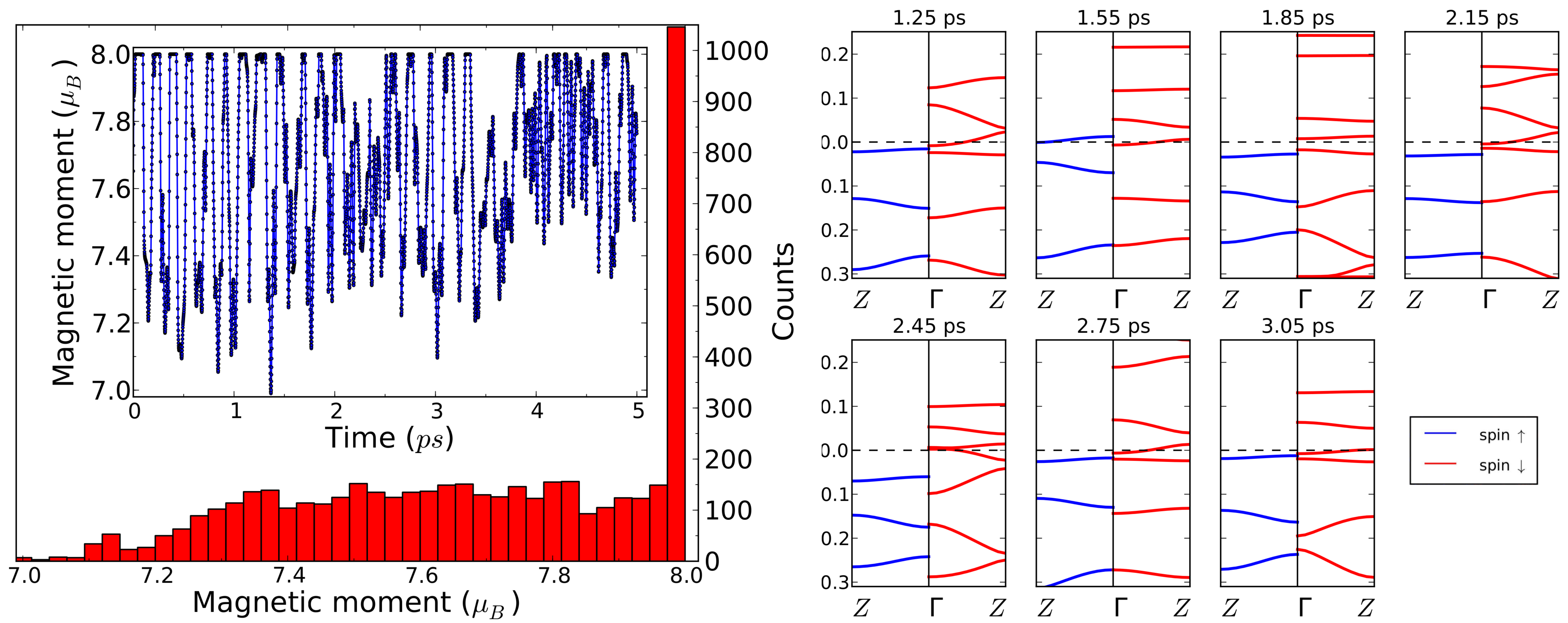}
 \caption{Inset of left panel shows the time evolution of the magnetic moment of a 
[AlN]$_6$(8) ribbon at room temperature. The ribbon adopts an 
undulating shape that slightly modifies the localized magnetic 
moment compared to 0 K value. The maximum magnetic moment 
deviation is $\sim$ -1 $\mu_B$/supercell. In the main panel the histogram 
shows the number of times that the system visits a magnetic
moment value along a 5 ps molecular simulation run where $\sim$1/5 
of the time the system has the maximum value of $\sim$2 $\mu_B$/unit cell. The right panels show the electronic band structure of a 8\% strained [AlN]$_6$(4) nanoribbon at several snapshots in time as indicated above each panel captured from a molecular dynamics run at T = 300 K. }                                                                                                                                                                                                                                                                                                                                             
 \label{fig5}
\end{figure*}

In that perspective, a very appealing strategy to 
tune the intrinsic half-metallicity of AlN nanoribbons 
is a combined approach based on the application of 
an external electric field to a strained ribbon. Our 
calculations predict that a +4\% strained [AlN]$_6$(2)
ribbon exhibits enhanced half-metallicity for E = 
-100 meV/\AA, and transitions into a semiconductor for 
E = +100 meV/\AA. Hence, while uniaxial strain allows 
for the enhancement of half-metallicity by controlling 
the magnetization density at the Fermi level, anexternal field allows us to reversibly tune the overall 
electron density at the Fermi level, driving the system 
either into a metallic or an insulating phase. 
It is worth pointing out that these transitions 
are intimately related to the spin arrangement at 
both edges and that for a spin alignment other than 
that presented above, the opposite phenomena are 
observed. We consider now the [AlN]$_6$(2) ribbon with 
a ferromagnetic spin distribution at both Al and N 
edges ($\uparrow\uparrow $-$\uparrow\uparrow$). This particular configuration might be 
obtained by applying a magnetic field perpendicular
to the ribbon surface. The metallic state observed in 
Figure \ref{fig2}(c) remains metallic for an external electric field 
of E = -100 meV/\AA\  and becomes semiconducting for 
E = +150 meV/\AA. The latter transition is accompanied
by an increase of the spin-up charge density at both 
the Al and N edge atoms, a decrease of the spin-down 
component at the N edge atoms, and no variation 
of it on the Al edge atoms. The magnetic moment 
saturates at 4.0 $\mu_B$ per unit cell. For the negative value 
of the applied electric field, spin-up components of 
both Al and N edge atoms gain some charge, the N 
edge atom releases some spin-down charge and the 
corresponding for Al remains the same. The magnetic 
moment decreases to 3.5 $\mu_B$ per unit cell. Therefore, 
unlike the $\uparrow\downarrow $-$\uparrow\uparrow$ spin distribution, in the case of the 
$\uparrow\uparrow $-$\uparrow\uparrow$ configuration, uniaxial strain barely affects the 
electronic bands responsible for the metallic character, 
and the spin-resolved self-doping is driven by the 
externally applied electric field. It does not induce 
half-metallicity due to the symmetric character of the 
magnetic polarization across the ribbon, but opens a 
band gap, rendering the ribbon semiconducting.

Additional calculations for the [AlN]$_{18}$(2) ribbons 
have been conducted to verify that the tunability is 
conserved for wider ribbons. Indeed by compressing 
the unit cell down to 2\% of the original length the 
half-metallic character of the nanoribbon is enhanced 
by lowering of the energy for one of the spin-channel
bands. Stretching renders the ribbon metallic, unlike 
the narrow ribbon that conserved the half-metallic 
behavior for larger unit cell parameters. The dependency of the electronic properties on the ribbon width 
clearly suggests the possibilities to explore different 
conducting or metallic regimes. Finally, applying a 
ribbon in-plane electric field perpendicular to the
axis modifies the [AlN]$_{18}$(2) ribbon electronic behavior 
in a similar manner to the [AlN]$_6$(2), namely a band 
gap develops as the applied field changes from 
negative to positive values (see Figure \ref{fig4}) .

To verify that the half-metallicity and the edge 
magnetic moments are robust at room temperature, 
we performed spin-polarized DFT-based molecular 
dynamics simulations with a Nos\'e thermostat at T = 
300 K on a segment of AlN ribbon 4-unit cell long, i.e., 
[AlN]$_6$(8) at 0\% and 8\% strain [18, 19]. The absence of 
large fluctuations of the total magnetic moment (inset of left panel of Figure \ref{fig5} for 0\%-strain and T = 300 K) is due to the highly
localized character of the spins on the ribbon edges
and the lack of orbital rearrangement during the ribbon
motion, which could otherwise lead to a quenching 
of magnetic moments. The half-metallicity is robust 
albeit with temperature dependent fluctuation of the 
electronic states (see the right panel of Figure \ref{fig5}).

\section{Conclusions}

We have computed the electronic and magnetic 
properties of zigzagged bare-edge AlN nanoribbons
under external traverse electric field and uniaxial strain.For the ground state, we predict that tensile strain 
selectively opens a band gap of one spin-channel, 
enhancing the half-metallicity, while compressive 
strain yields semiconducting behavior. Depending on 
strength and direction, transverse electric fields tune 
the electronic structure from half-metallic to metallic 
or semiconducting behavior. The applied fields induce 
a charge polarization across the ribbon with no spin 
selectivity. In contrast, uniaxial tensile strain induces 
a selective spin-resolved charge screening which 
self-dopes the $\pi$-orbitals of the Al-edge atoms due to 
a net charge transfer from the opposite N-edge atoms. 
For certain magnetic states lying energetically close 
to the ground state, opposite behaviors are found. 
The ribbon remains half-metallic even at finite temperatures. To our knowledge, this AlN nanoribbon 
tunability has not been observed or predicted in 
other nanoribbons and has its origin at edge states
derived from orbitals with sufficiently different electron
localization. 
 
The presence of opposite local magnetic densities 
on the Al-edge easily leads to a tunable spin-resolved
doping of the more metallic Al-edge from the more 
localized N-edge states. Since the electronic control is
dependent on the effective screening between the
two edges in the nanoribbon, increasing the distance 
between the edges is expected to reduce the 
tunability of the system under strain. The tunability 
is significant at least up to a thickness of 18-AlN units 
in our simulations. The predicted wide tunability of 
the electronic structure should be independent of the 
electronic structure theory.

\section{Acknowledgments}

This research used resources of the National Center 
for Computational Sciences at Oak Ridge National 
Laboratory (ORNL), under Contract No. DE-AC05-
00OR22725 and the National Energy Research 
Scientific Computing Center, under Contract No. 
DE-AC02-05CH11231, both supported by the Office 
of Science of the U.S. DOE. We acknowledge support 
from the Center for Nanophase Materials Sciences
(CNMS), sponsored at ORNL by the Division of
Scientific User Facilities, U.S. Department of Energy.

\section{Bibiliography}

[1] Wolf, S. A.; Awschalom, D. D.; Buhrman, R. A.; Daughton, 
J. M.; von Moln\'ar, S.; Roukes, M. L.; Chtchelkanova, A. Y.; 
Treger, D. M. Spintronics: A spin-based electronics vision 
for the future. Science 2001, 294, 1488-1495. 

[2] Heusler, O. Kristallstruktur und ferromagnetismus der 
mangan-aluminium-kupferlegierungen. Ann. Phys. 1934, 
411, 155-201. 

[3] de Groot, R. A.; Mueller, F. M.; vanEngen, P. G.; Buschow, 
K. H. J. New class of materials: Half-metallic ferromagnets. 
Phys. Rev. Lett. 1983, 50, 2024-2027.

[4] Dolui, K.; Pemmaraju, C. D.; Sanvito, S. Electric field effects 
on armchair MoS$_2$ nanoribbons. ACS Nano 2012, 6, 4823-4834. 

[5] Du, A. J.; Zhu, Z. H.; Chen, Y.; Lu, G. Q.; Smith, S. C. First 
principle studies of zigzag AlN nanoribbon. Chem. Phys. 
Lett. 2009, 469, 183-185. 

[6] Son, Y.-W.; Cohen, M. L.; Louie, S. G. Half-metallic 
graphene nanoribbons. Nature 2006, 444, 347-349. 

[7] Hod, O.; Barone, V.; Peralta, J. E.; Scuseria, G. E. Enhanced 
half-metallicity in edge-oxidized zigzag graphene nanoribbons. 
Nano Lett. 2007, 7, 2295-2299.

[8] Qi, J. S.; Qian, X. F.; Qi, L.; Feng, J.; Shi, D. N.; Li, J. Strain- 
engineering of band gaps in piezoelectric boron nitride 
nanoribbons. Nano Lett. 2012, 12, 1224-1228. 

[9] Piazza, G.; Felmetsger, V.; Muralt, P.; Olsson III, R. H.; 
Ruby, R. Piezoelectric aluminum nitride thin films for 
microelectromechanical systems. MRS Bull. 2012, 37, 1051- 
1061. 

[10] O'Leary, S. K.; Foutz, B. E.; Shur, M. S.; Eastman, L. F. 
Steady-state and transient electron transport within the III-V 
nitride semiconductors, GaN, AlN, and InN: A review. J. 
Mater. Sci.: Mater. El. 2006, 17, 87-126. 

[11] Balasubramanian, C.; Bellucci, S.; Castrucci, P.; De Crescenzi, 
M.; Bhoraskar, S. V. Scanning tunneling microscopy 
observation of coiled aluminum nitride nanotubes. Chem. 
Phys. Lett. 2004, 383, 188-191. 

[12] Zhukovskii, Y. F.; Popov, A. I. Balasubramanian, C.; Bellucci, 
S. J. Phys.: Condens. Matter 2006, 18, 2045. 

[13] Xie, T.; Lin, Y.; Wu, G. S.; Yuan, X. Y.; Jiang, Z.; Ye, C. 
H.; Meng, G. W.; Zhang, L. D. AlN serrated nanoribbons
synthesized by chloride assisted vapor-solid route. Inorg. 
Chem. Commun. 2004, 7, 545-547. 

[14] Moon, E. J.; Rondinelli, J. M.; Prasai, N.; Gray, B. A.; 
Kareev, M.; Chakhalian, J.; Cohn, J. L. Strain-controlled 
band engineering and self-doping in ultrathin LaNiO3 films. 
Phys. Rev. B 2012, 85, 121106-121109. 

[15] Ordej\'on, P.; Artacho, E.; Soler, J. M. Self-consistent order-N 
densisty-functional calculations for very large systems. Phys. 
Rev. B 1996, 53, 10441-10444.

[16] Soler, J. M.; Artacho, E.; Gale, J. D.; Garc\'ia, A.; Junquera, 
J.; Ordej\'on, P.; S\'anchez-Portal, D. The SIESTA method for 
ab initio order-N materials simulation. J. Phys.: Condens. 
Matter 2002, 14, 2745-2779. 

[17] Clement, M.; Vergara, L.; Sangrador, J.; Iborra, E.; Sanz- 
Herv\'as, A. SAW characteristics of AlN films sputtered on 
silicon substrates. Ultrasonics 2004, 42, 403-407. 

[18]  Kresse G.; Furthm\"uller, J. Efficient iterative scheme for ab
initio total-energy calculations using a plane-wave basis set. 
Phys. Rev. B 1996, 54, 11169-11186. 

[19] Kresse G.; Joubert, D. From ultrasoft pseudopotentials tothe projector augmented-wave method. Phys. Rev. B 1999, 
59, 1758-1775.

[20] Liu, Q. Q.; Yu, X. H.; Wang, X. C.; Deng, Z.; Lv, Y. X.; 
Zhu, J. L.; Zhang, S. J.; Liu, H. Z.; Yang, W. G.; Wang, L.; 
et al. Pressure-induced isostructural phase transition and 
correlation of FeAs coordination with the superconducting
properties of 111-type Na1-xFeAs. J. Am. Chem. Soc. 2011, 
133, 7892-7896.

\end{document}